\documentclass[twocolumn,superscriptaddress,amsmath,amssymb,showpacs,showkeys]{revtex4}

\usepackage{graphicx,color}
\usepackage{dcolumn}
\usepackage{bm}
\newcommand{\beq}{\begin{equation}} \newcommand{\eeq}{\end{equation}}
\newcommand{\bqa}{\begin{eqnarray}} \newcommand{\eqa}{\end{eqnarray}}

\definecolor{gold}{rgb}{0.75,0.56,0.00}
\definecolor{green}{rgb}{0.00,0.50,0.00}

\newcommand{\ms}[1]{\mbox{\scriptsize #1}}

\begin{document}


\title{Observing quantum chaos with noisy measurements and highly mixed states}

\author{Jason F. Ralph}
 \email{jfralph@liverpool.ac.uk}
 \affiliation{Department of Electrical Engineering and Electronics, University of Liverpool,  Brownlow Hill, Liverpool, L69 3GJ, UK.}
\author{Kurt Jacobs}
 \email{kurt.jacobs@umb.edu}
  \affiliation{U.S. Army Research Laboratory, Computational and Information Sciences Directorate, Adelphi, Maryland 20783, USA.}
  \affiliation{Department of Physics, University of Massachusetts at Boston, Boston, MA 02125, USA} 
\affiliation{Hearne Institute for Theoretical Physics, Louisiana State University, Baton Rouge, LA 70803, USA} 
\author{Mark J. Everitt}
 \email{m.j.everitt@lboro.ac.uk}
 \affiliation{Quantum Systems Engineering Research Group, Department of Physics, Loughborough University,  Loughborough, LE11 3TU, UK.}

\date{\today}

\begin{abstract}
A fundamental requirement for the emergence of classical behavior from an underlying quantum description is that certain observed quantum systems make a transition to chaotic dynamics as their action is increased relative to $\hbar$. While experiments have demonstrated some aspects of this transition, the emergence of quantum trajectories with a positive Lyapunov exponent has never been observed directly. Here, we remove a major obstacle to achieving this goal by showing that, for the Duffing oscillator, the transition to a positive Lyapunov exponent can be resolved clearly from observed trajectories even with measurement efficiencies as low as 20\%. We also find that the positive Lyapunov exponent is robust to highly mixed, low purity states and to variations in the parameters of the system. 
\end{abstract}

\pacs{05.45.Mt, 03.65.Ta, 05.45.Pq}
\keywords{quantum state-estimation, continuous measurement, quantum chaos, classical limit}

\maketitle


The emergence of classical chaotic-like behaviour from quantum mechanical systems has been an area of active research for many years~\cite{Haa2013}. There has been a great deal of interest in purely quantum systems, displaying unitary evolution, and non-unitary open quantum systems. This paper is concerned with open quantum systems whose classical counterparts are chaotic and make a transition to chaotic behavior as their size (more precisely their action) is increased so as to be large compared to $\hbar$~\cite{Spi1994,Sch1995,Zur1995,Bru1996,Bru1997,Hab1998,Bha2000,Mil2000,Ste2001,Sco2001,Bha2003,Eve2005a,Eve2005b,Sha2013,Len2013,Eas2016,Pok2016,Nei2016}. This transition is enabled by their interaction with the environment or when they are subjected to continuous observation. In the former case, the evolution approaches that of the probability density in phase space for the equivalent classical system as the action is increased~\cite{Zur1995,Hab1998,Sha2013}. Continuous observation turns this probability density into individual trajectories that follow the nonlinear classical motion with the requisite Lyapunov exponents~\cite{Bha2003,Sha2013,Eas2016,Pok2016}. 
 
Recent experimental progress in the control and measurement of quantum systems has enabled the continuous measurement of individual quantum systems and the calculation of quantum trajectories and state estimates~\cite{Mur2013,Web2014,Six2015,Cam2016}. This opens up the exciting possibility of directly observing the trajectories of classical chaotic dynamics emerging in open quantum systems. By observing a sufficiently long trajectory, it should also be possible to identify positive Lyapunov exponents, as a fundamental characteristic parameter that is indicative of chaos. Although experiments have been performed to explore the quantum-classical transition~\cite{Mil2000,Ste2001,Len2013,Nei2016} and to identify aspects of chaotic behavior in open quantum systems, the Lyapunov exponents have not been determined experimentally from quantum trajectories. One of the difficulties in such experiments is the efficiency of the measurement process. In an ideal measurement, the noise will be purely quantum in origin and the measurement efficiency, defined to be the fraction of the noise power due to the quantum measurement as opposed to extraneous classical noise from other sources, will be 100\%. Unfortunately, practical measurement systems are often far from ideal, and even the best experiments have efficiencies well below 100\%. For example, the experimental efficiencies reported in \cite{Web2014} are around 35\%. For the observation of certain purely quantum effects, the efficiency must be above some minimum threshold level. {\em Rapid-purification}~\cite{Jac2003, Com2006, Wis2006, Hil2007, Com2008, Wis2008, Li2013}, for example, requires a measurement efficiency of at least 50\% \cite{Li2013}. 

In this paper, we show that a positive Lyapunov exponent and the associated transition to classical chaos could be derived from quantum trajectories and continuous measurements with efficiencies as low as 20\%. Further, we find that the value of the positive Lyapunov exponent is robust across a wide range of purities, and are insensitive to variations in system parameters of at least $5\%$. This opens the way to observing the emergence of chaos in open quantum systems with current technology. 

The evolution of a continuously observed quantum system is described by a stochastic master equation~\cite{Bel1999,Wis2010,Jac2014}. As such, our work here is aided greatly by a recent and significant improvement in the numerical methods available to solve such equations, due to Rouchon and collaborators~\cite{Ami2011,Rou2015}. It also benefits from the ``moving basis'' method used by Schack, Brun and Percival~\cite{Sch1995, Bru1996}. The system that we consider is a standard example from classical chaos: the Duffing oscillator. This system has been studied for pure states and efficient measurements and it has been shown to make a transition from non-chaotic to chaotic motion as the action is increased relative to $\hbar$~\cite{Bru1996,Bru1997,Bha2000,Sco2001,Bha2003,Eve2005a,Eve2005b,Eas2016,Pok2016}. To achieve this, one must change the mass of the oscillator, the potential, and any driving forces in such a way that the dynamics remain the same up to a scaling of the coordinates and time, while the area of the phase space increases with respect to $\hbar$. A simple way to do this is to first write the Hamiltonian of the system, $\hat{H}$, in terms of dimensionless variables $\hat{q}$ and $\hat{p}$, then to change the size of the phase space by defining the new Hamiltonian to be $\hat{H}_{\beta} = \beta^{-2} \hat{H}(\beta \hat{q}, \beta \hat{p})$. The overall factor of $\beta^{-2}$ merely scales time. The size of the phase space for the Hamiltonian $\hat{H}_{\beta}$ now scales as $\beta^{-2}$ so that the classical limit is given by $\beta \rightarrow 0$~\cite{Sch1995,Bru1996}. 

The resulting dimensionless Hamiltonian for the Duffing oscillator is 
\begin{equation}\label{Ham}
\hat{H}_\beta = \frac{1}{2}\hat{p}^2+\frac{\beta^2}{4}\hat{q}^4-\frac{1}{2}\hat{q}^2+\frac{g}{\beta}\cos(t)\hat{q}+\frac{\Gamma}{2}(\hat{q}\hat{p}+\hat{p}\hat{q}) .  
\end{equation} 
The first term in $\hat{H}_\beta$ is the kinetic energy, the second and third terms give the double-well potential, and the fourth is the periodic linear driving with a tunable amplitude $g/\beta$. The final term in the expression for $\hat{H}_\beta$ may look unusual, and is included because, in combination with the dissipative measurement process, it generates linear damping in momentum. (The Markovian dissipative measurement damps both $\hat{p}$ and $\hat{q}$. The Hamiltonian term amplifies $\hat{q}$ and damps $\hat{p}$, thus canceling the damping of $\hat{q}$ so that only the damping of $\hat{p}$ remains \cite{Duf2016}). While damping of momentum is not required to observe chaos in the Duffing oscillator~\cite{Bha2000}, it is useful in numerical simulations to keep the phase space contained. In terms of the real physical position $\hat{X}$, the momentum $\hat{P}$, and the Hamiltonian $\hat{H}_{\ms{r}}$, the scaled variables are given by $\hat{q} = \hat{X}/\sqrt{\hbar/m\omega}$, $\hat{p} = \hat{P}/\sqrt{\hbar m \omega}$, and $\hat{H}_\beta = \hat{H}_{\ms{r}}/(\hbar\omega)$, in which $m$ is the mass of the oscillator and $\omega$ is an arbitrary frequency scale. 

Since the observer will not have full information about which pure state the system is in at any given time, the observer's knowledge about this state is described by the density matrix, $\rho$. The purity of the density matrix is defined by $P = \mathrm{Tr}[\rho^2]$, and indicates the level of certainty that the observer has about the system's state. Under the action of a continuous measurement, the evolution of the density matrix is stochastic. This is due to the fact that the stream of measurement results necessarily has a fluctuating component, and the density matrix is a full description of the  
observerÕs state-of-knowledge conditioned on these results. To emphasize this, we will denote the density matrix by $\rho_{\ms{c}}$. 

For the continuous measurement, we use a standard model in which a transmission line --- or more generally a medium that supports a continuum of traveling waves --- is coupled to the system so as to mediate both damping and the extraction of information~\cite{Wis2010,Jac2014}. The exact type of measurement has been shown to be unimportant in observing the emergence of classical dynamics, so long as it provides enough information about the position and momentum to maintain sufficient localization of the state in phase space~\cite{Bha2000, Sch1995, Bru1996}. In fact, for the work presented here, we have also performed the simulations using a continuous measurement of the position, $\hat{q}$, and this showed very similar behavior.

Under the action of continuous measurement, the evolution of the density matrix is given by the stochastic master equation (SME)~\cite{Wis2010,Jac2014},
\begin{eqnarray}\label{sme1}
d\rho_{\ms{c}}&=&- i \left[\hat{H}_\beta,\rho_{\ms{c}}\right]dt \nonumber \\
&&+\left\{ \hat{L} \rho_{\ms{c}} \hat{L}^{\dagger} -\frac{1}{2}\left(\hat{L}^{\dagger} \hat{L} \rho_{\ms{c}} + \rho_{\ms{c}} \hat{L}^{\dagger} \hat{L} \right)\right\}dt   \nonumber \\
&&+\sqrt{\eta}\left(\hat{L}\rho_{\ms{c}}+\rho_{\ms{c}} \hat{L}^{\dagger}-\mathrm{Tr}[\hat{L}\rho_{\ms{c}}+\rho_{\ms{c}} \hat{L}^{\dagger}] \right)dW \nonumber \\
\end{eqnarray}
in which $\hat{L} = \sqrt{2\Gamma}\hat{a}$, with $\hat{a} = (\hat{q} + i \hat{p})/\sqrt{2}$, and the stream of measurement results (the ``measurement record'') is given by   
\begin{equation}\label{record}
   y(t+dt) =  y(t) + \sqrt{\eta}\mathrm{Tr}[\hat{L}\rho_{\ms{c}}+\rho_{\ms{c}} \hat{L}^{\dagger}] dt+dW  
\end{equation}
where $dW$ are increments of Weiner noise and thus satisfy $\langle dW \rangle =0$ and $dW^2 = dt$. The efficiency of the measurement is denoted by $\eta$, and is defined to be the fraction of the noise power due to the measurement rather than other (classical) sources of noise, i.e. the fraction of the output signal that is recorded by the measuring device. 

For Rouchon's method~\cite{Ami2011,Rou2015} with a moving basis, the increment to $\rho_{\ms{c}}$ for the time step from $t_n = n\Delta t$ to $t_{n+1} = (n+1)\Delta t$, is given by $\Delta \rho_{\ms{c}}^{(n)} = \rho_{\ms{c}}^{(n+1)}- \rho_{\ms{c}}^{(n)}$, where
\begin{equation}\label{sme2}
\rho_{\ms{c}}^{(n+1)}= \frac{\hat{M}_n\rho_{\ms{c}}^{(n)}\hat{M}_n^{\dagger}+(1-\eta)\hat{L}\rho_{\ms{c}}^{(n)}\hat{L}^{\dagger}\Delta t }
{\mathrm{Tr}\left[\hat{M}_n\rho_{\ms{c}}^{(n)}\hat{M}_n^{\dagger}+ (1-\eta)\hat{L}\rho_{\ms{c}}^{(n)}\hat{L}^{\dagger}\Delta t \right]}
\end{equation}
and $\hat{M}_n$ is given by 
\begin{eqnarray}\label{Mn1}
\hat{M}_n &=& I-\left(i\hat{H} +\frac{1}{2} \hat{L}^{\dagger}\hat{L}\right)\Delta t +\frac{\eta}{2}\hat{L}^2(\Delta W(n)^2-\Delta t)  \nonumber\\
&& +\sqrt{\eta}\hat{L}\left(\sqrt{\eta}\mathrm{Tr}[\hat{L}\rho_{\ms{c}}^{(n)}+\rho_{\ms{c}}^{(n)}\hat{L}^{\dagger}]\Delta t +\Delta W(n)\right), 
\end{eqnarray}
where the $\Delta W$'s are independent Gaussian variables with zero mean and a variance equal to $\Delta t$. To represent the density matrix we use a harmonic oscillator (Fock) basis, changing the location of this oscillator to follow the expected location of the system in phase space (i.e. the expectation values of $\hat{q}$ and $\hat{p}$). This greatly reduces the size of the state-space required for the simulation. Figure 1 shows an example trajectory in the chaotic regime.   

To verify whether a system exhibits chaotic behavior or not, it is necessary to calculate the Lyapunov exponents for the trajectory. In classical dynamics, this is fairly straightforward and uses the Jacobian, calculated from the classical dynamical equations, and the Lyapunov exponents are found from the eigenvalues of the product of the Jacobian matrices along the trajectory and taking the infinite time limit \cite{Sko2010}. In practice, the Lyapunov exponents are estimated in the long (but finite) time limit and the Jacobian products are repeatedly renormalized using a Gram-Schmidt orthonormalization procedure to constrain the tendency of the eigenvalues to increase beyond the numerical limits of the computer \cite{Sko2010}. For quantum systems, a number of approaches have been proposed and used to define Lyapunov exponents~\cite{Haa1992, Zyc1993,Jal2001}. Here, the generation of trajectories means that an approach analogous to the classical calculation method can be used~\cite{Jal2001}, but rather than using the classical dynamical equations to generate a Jacobian at each time step, an approximate Jacobian is constructed using the evolution of the expectation values for $\hat{q}$ and $\hat{p}$ under the non-stochastic evolution given by (\ref{sme2}), i.e. the evolution predicted when $dW=0$. Because of these factors, the finite time of the simulation and the differences in the construction of the Jacobian matrices, the solutions that generate a positive Lyapunov exponent are strictly chaotic-like rather than true chaos in the mathematical sense. However, we refer to the solutions as chaotic for reasons of practicality. 

With a two-dimensional phase space and an arbitrary phase variable for the drive term, we would expect to obtain three Lyapunov exponents, one of which would always be zero (corresponding to perturbations along the trajectory). We will denote the two non-zero Lyapunov exponents by $\lambda_{+}$ and $\lambda_{-}$ respectively, noting that $\lambda_{+}$ could be positive (chaotic solution) or negative (periodic solution) and $\lambda_{+}+\lambda_{-} <0$. The estimates of the Lyapunov exponents calculated below were obtained using the parameter values $g=0.3$ and $\Gamma=0.125$, with a moving basis containing between 80 and 200 oscillator states, and between 2000 and 6000 time increments per cycle of the drive term. The size of the basis and time steps was varied to ensure that the integration of the SME was numerically stable. Although the values of the Lyapunov exponents are found to be insensitive to measurement inefficiencies, the state estimates generated using (\ref{sme1}) and a particular measurement record (\ref{record}) can be numerically unstable if the basis contains insufficient numbers of states or the time increments are too large. For measurement efficiencies around 20\% and $\beta$ values around $0.1$, the number of states required to generate a stable trajectory in the moving basis grows to 150-200 states and the time step must be $\Delta t \simeq \pi/3000$.  
\begin{figure}[t] 
   \centering
   \includegraphics[width=1\hsize]{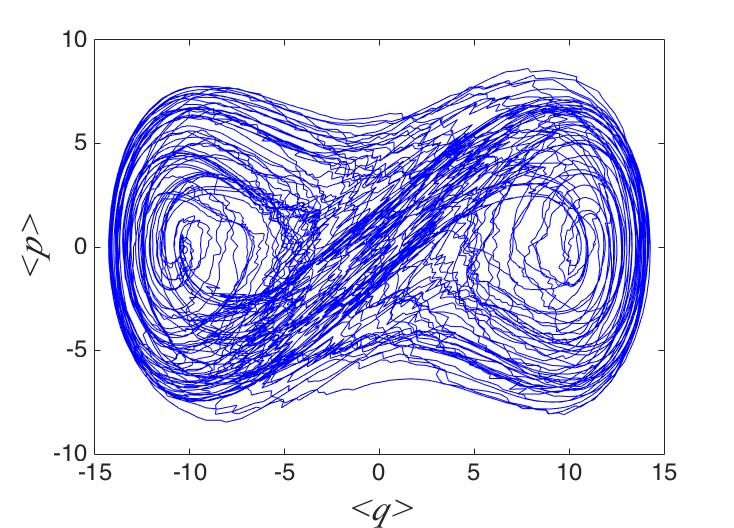} 
   \caption{An example quantum trajectory for $\beta = 0.1$ and $\eta = 0.4$, with $g=0.3$ and $\Gamma=0.125$.}
   \label{fig:phaseplot}
\end{figure}

\begin{figure}[t] 
   \centering
   \includegraphics[width=1\hsize]{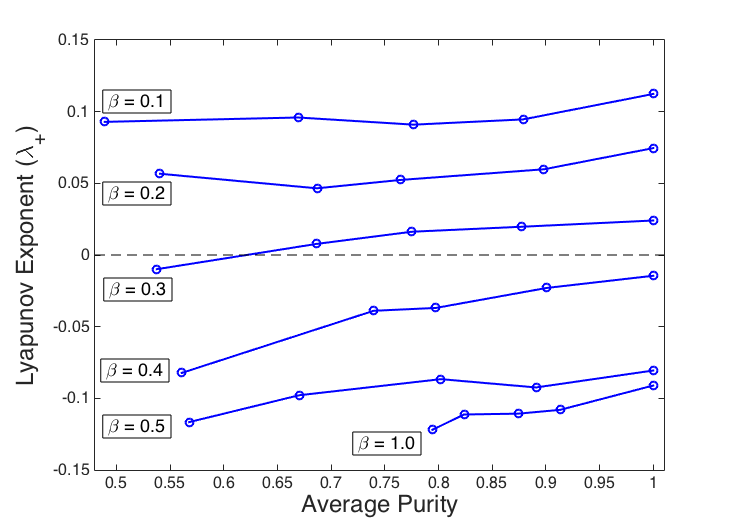} 
   \caption{The largest non-zero Lyapunov exponent ($\lambda_{+}$) calculated for $\beta=1.0 - 0.1$ as a function of the average purity of the estimated state after 100 cycles of the drive term, $t=200\pi$, with $g=0.3$ and $\Gamma=0.125$ (solid lines). The points marked correspond to measurement efficiencies of $\eta = 1.0$, $0.8$, $0.6$, $0.4$ and $0.2$ (right to left).}
   \label{fig:lyapunov1}
\end{figure}

Figure~\ref{fig:lyapunov1} shows the largest non-zero Lyapunov exponents estimated for $\beta$ values between 1.0 (noisy-periodic) and 0.1 (noisy-chaotic) for measurement efficiencies from 20\% to 100\%. A small number of simulations were performed for measurement efficiencies as low as 10\%. It was possible to obtain values for positive exponents in some cases but the numerical stability of the SME was affected for $\beta < 0.2$ so these results are not shown. The main feature to note in Figure~\ref{fig:lyapunov1} is that the positive Lyapunov exponents are approximately constant as a function of purity and for the range of measurement efficiencies, up to some small fluctuations due to the stochastic nature of the trajectories. There is a weak linear dependence on the average purity for the negative exponents (noisy-periodic trajectories). The figure also shows that the periodic solutions often have a higher average purity for the same measurement efficiency. The chaotic solutions have lower purities except for cases where $\eta = 1.0$, which will always asymptote to a pure state $P=1$, because all of the information contained in the measurement record is available to construct the quantum state. This relationship between periodic solutions and a higher average purity might be expected intuitively and has been noted in\cite{Sha2013}. Chaos is associated with information ``creation'', in that two chaotic solutions from neighboring points will diverge as the small differences are amplified by the stretching and folding of phase space associated with chaotic evolution \cite{Sko2010} -- although not shown, this stretching and folding process can be seen in the quantum states if the phase space Wigner functions are plotted on the $q-p$ plane \cite{Sha2013}. As a result of this, it could be anticipated that a chaotic trajectory with a positive Lyapunov exponent would require more measurements to extract the information required to construct an accurate state estimate, and an inefficient measurement would be likely to produce a less accurate state estimate for chaotic evolution than for periodic evolution. The minimum Lyapunov exponents ($\lambda_{-}$) are all negative, as expected. They are not shown explicitly, but they were also found to be relatively insensitive to the purity of the states and the efficiency of the measurements. 

\begin{figure}[t] 
   \centering
   \includegraphics[width=1\hsize]{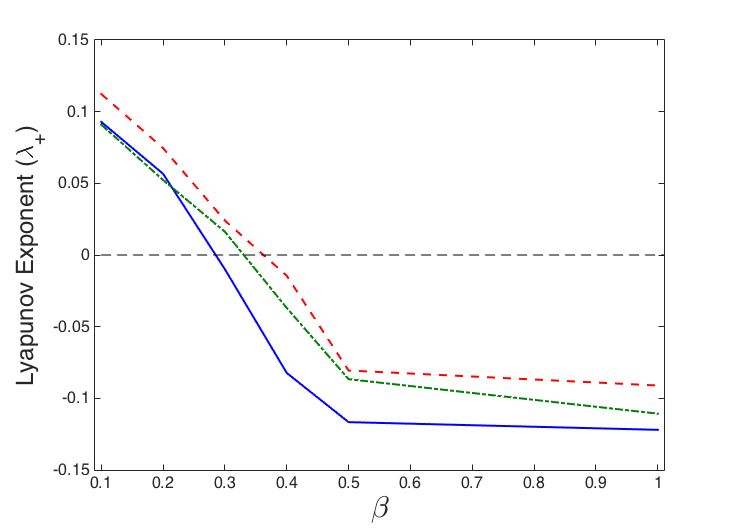} 
   \caption{The largest non-zero Lyapunov exponents ($\lambda_{+}$) calculated for $\beta= 1.0-0.1$ for measurement efficiencies of $\eta = 1.0$ (red-dashed line), $0.6$ (dot-dashed-green line) and $0.2$ (blue-solid line), after 100 cycles of the drive term, $t=200\pi$, with $g=0.3$ and $\Gamma=0.125$.}
   \label{fig:lyapunov2}
\end{figure}

Figure \ref{fig:lyapunov2} shows the largest non-zero Lyapunov exponents as functions of $\beta$, as in \cite{Eas2016}, for three different measurement efficiencies. These represent the transition from the quantum ($\beta = 1.0$) to the near classical regime ($\beta=0.1$). Positive Lyapunov exponents and chaotic behavior appear at $\beta = 0.3$ \cite{Eas2016}. The figure shows that the transition is preserved even when the measurement efficiency and, consequently, the average purity of the quantum states are low, which is relevant for possible experimental investigations where the measurements are not idealized theoretical models.

The accuracy of the estimation process and of the quantum trajectory are reliant on the accuracy of the Hamiltonian and the parameters used in the SME to estimate the quantum state from the measurement record. If there is a mis-match between the system that generates the measurements and the parameter values used in the SME, the fidelity of the quantum state estimate will be adversely affected. It is natural, therefore, to ask what effect such mis-matches would have on the estimation of the Lyapunov exponents. To address this concern, simulations were conducted using one filter to generate a continuous measurement record, and this record was then fed into a second SME, where the second SME had errors in each of the parameters in the Hamiltonian (\ref{Ham}) and the SME (\ref{sme1}): $g$, $\beta$, $\Gamma$, $\eta$, and the initial phase of the cosine drive term. In each case, the accuracies of the quantum trajectories did deteriorate, but the estimates for the Lyapunov exponents were found to be insensitive to errors up to 5\% of the true parameter values. So, the Lyapunov exponents were found to be robust against measurement inefficiencies, highly mixed states and mis-matches in the state estimation processing. 

The importance of these results lies in the accessibility of the characteristic Lyapunov exponents to experimental investigation. As we have already noted, continuous measurements are difficult to achieve in experiments and are often limited in terms of their efficiency \cite{Web2014}. A signature of chaos that is related to the ``quantum-ness'' or classicality of the system and that is relatively insensitive to the measurement efficiency could be a significant factor in the experimental observation of quantum chaos in such systems. The signature is also robust against highly mixed states and inaccuracies in the experimental parameters. It is also a benefit that the Duffing oscillator can be realized using superconducting circuits and Josepson junctions \cite{Man2007,Guo2010} and it already forms the basis for nonlinear amplifiers used in quantum circuit experiments \cite{Man2007}. Using the notation given in \cite{Man2007}, we can define dimensionless quantities for the parallel circuit configuration (also called the radio-frequency SQUID \cite{Bar1983}): $\hat{q}=\Phi (\hbar\sqrt{L_p/C_p})^{-\frac{1}{2}}$, $\hat{p}=Q (\hbar\sqrt{C_p/L_p})^{-\frac{1}{2}}$, where $\Phi$ is the magnetic flux, $Q=C_p\dot{\Phi}$ is the conjugate momentum, $C_p$ is the junction capacitance, $L_p$ is the parallel inductance formed from the Josephson inductance $L_J$ and the geometric inductance $L_{pe}$ and $\omega=1/\sqrt{C_p L_p}$. To produce the potential given in (\ref{Ham}), the circuit must be biased to give a negative quadratic term and a positive quartic term, with $L_p=(1/L_J-1/L_{pe})^{-1}$. For this configuration, the classical scaling parameter is given by $\beta=\sqrt{e/(3\omega C_p(1-L_J/L_{pe}))}$, where $e$ is the electron charge, and the classical limit is taken by letting the effective ``mass'' of the system $C_p\rightarrow\infty$.

In this paper, we have studied the properties of the quantum Duffing oscillator in the presence of a continuous measurement, mediated by a weak coupling to an environment. The stochastic master equation was used to follow the evolution of the quantum state, for both ideal (efficient) measurements and inefficient measurements; including very inefficient measurements, leading to highly mixed states. The resultant quantum trajectories are stochastic and can exhibit periodic or chaotic behavior as the dynamical evolution is scaled from the quantum regime towards the classical limit. The standard indicators of chaos, the Lyapunov exponents, have been calculated for this system. Positive Lyapunov exponents were shown to be insensitive to the measurement efficiency and to the purity of the quantum states, meaning that the emergence of chaotic behavior can be determined even when using very inefficient measurements and highly mixed states. The Lyapunov exponents calculated from the quantum trajectories were also found to be robust to variations in all of the parameter values used in the state estimation process. The robustness of the Lyapunov exponents to these factors would be significant for any experimental investigation of chaos in open quantum systems, because it demonstrates that the quantum-classical transition to chaotic behavior should be accessible even when the measurements are not ideal and the system parameters have not been characterized perfectly.

\textit{Acknowledgments:} JFR would like to thank the US Army Research Laboratories (contract no. W911NF-16-2-0067).

\bibliographystyle{apsrev}

\end{document}